\documentclass[a4paper,aps,prd,10pt,preprintnumbers,showpacs,superscriptaddress,nofootinbib,amsmath,amssymb,floatfix
,twocolumn
]{revtex4-1}

\usepackage{enumitem}
\usepackage{graphicx}
\usepackage[utf8]{inputenc}
\usepackage[T1]{fontenc}
\usepackage{cmap}
\usepackage{graphicx}
\usepackage{dcolumn}
\usepackage{bm}
\usepackage{bbm}
\usepackage{amsfonts}
\usepackage{amsmath}
\usepackage{enumitem}
\usepackage{lipsum}
\usepackage[bottom]{footmisc}
\usepackage{cmap}
\usepackage{subcaption}
\usepackage{braket}
\usepackage{braket}
\usepackage[table]{xcolor}
\usepackage{amsmath}
\usepackage{hyperref}
\usepackage{cleveref}
\usepackage{fmtcount}
\usepackage{dblfnote}
\usepackage{float}
\usepackage{amssymb}
\usepackage{natbib}
\usepackage{blindtext}

\begin{document}

\title{Universal topological classes of black holes surrounded by quintessence}

\author{Meng-Yao Zhang}
\email{myzhang94@yeah.net}
\affiliation{College of Big Date and Intelligent Engineering, Guizhou University of Commerce, Guiyang, 550014, China}
\author{Hou-You Zhou}
\email{hyzhou2021@yeah.net}
\affiliation{School of Mechanics and Civil Engineering, China University of Mining \& Technology, Beijing, Beijing 100083, China}
\author{Hao Chen}
\email{haochen1249@yeah.net}
\affiliation{School of Physics and Electronic Science, Zunyi Normal University, Zunyi 563006, China}
\author{Hassan Hassanabadi}
\email{hha1349@gmail.com}
\affiliation{Departamento de F\'{\i}sica Te\'orica, At\'omica y Optica and Laboratory for Disruptive \\ Interdisciplinary Science (LaDIS), Universidad de Valladolid, 47011 Valladolid, Spain}
\affiliation{Department   of   Physics, Faculty of Science,   University   of   Hradec   Kr\'{a}lov\'{e},  Rokitansk\'{e}ho 62, 500   03   Hradec   Kr\'{a}lov\'{e},   Czechia}
\author{Zheng-Wen Long}
\email{zwlong@gzu.edu.cn }
\affiliation{College of Physics, Guizhou University, Guiyang, 550025, China}

\begin{abstract}
This work explores the universal topological classes of various black hole configurations immersed in a quintessence field. The results indicate that the Schwarzschild black hole surrounded by quintessence is of the $W^{1-}$ type, exhibiting thermodynamic instability in both extremal temperature limits. The introduction of rotation alters its topological nature to the $W^{0+}$ class, where the Kerr black hole displays coexistence of a stable small black hole and an unstable large black hole at low temperatures. In comparison, anti de Sitter (adS) black holes display different thermodynamic topologies. The Schwarzschild–adS solution corresponds to the $W^{0-}$ class, where an unstable small branch and a stable large branch emerge in the high temperatures. Meanwhile, the Kerr–adS classified as $W^{1+}$, maintains stability in both high- and low-temperature limits. Furthermore, the 4D/5D quintessential black hole belongs to the $W^{0+}$ class within the Einstein–Gauss–Bonnet framework. Overall, parameters such as rotation and Gauss-Bonnet in quintessence background field affect the topological classifications of the black holes, while the quintessence field primarily modifies the stability range without changing the topological class itself.
\end{abstract}
\maketitle

\section{Introduction}\label{Sec1}
Black hole thermodynamics, as a crucial avenue for exploring the deep connections between black hole properties and thermodynamic principles, originates from a groundbreaking realization: black holes possess thermodynamic quantities such as entropy and temperature ~\cite{H1,H2,H3,H4}. Although substantial progress has been made in this field, a unified and systematic framework for characterizing the intrinsic features of various black hole systems is still lacking. In recent years, the emergence of the thermodynamic topology approach has provided a novel perspective. This method treats black holes as topological defects in thermodynamic space and defines topological invariants by analyzing the singularity structure of generalized free energy. Based on this framework, most black holes can be classified into three distinct topological classes: +1, 0, and –1, which reflect their global stability~\cite{Wei2}. Due to its generality and computational efficiency, this approach has been widely applied to black holes in diverse settings, including rotating AdS black holes~\cite{T4,T5}, phantom AdS black holes~\cite{chen}, Lovelock AdS black holes ~\cite{T6,T7}, Born-Infeld black holes ~\cite{T9}, as well as more complex configurations such as nonlinearly charged black holes ~\cite{T10}, R-charged black holes~\cite{T11} and hairy black holes ~\cite{T8}. Furthermore, topology has also been extended to the study of black holes in R\'{e}nyi statistics ~\cite{T12,T13}. Recently, Ref. ~\cite{T14} further refined the topological classification of black hole thermodynamics into four universal topological types. Subsequently, Ref. ~\cite{T15} expanded the existing classification framework. The new topological classification method has already been applied to other types of black holes ~\cite{T16,T17,T18,T19}. This topological framework opens new avenues for understanding the geometric nature of gravity. Naturally, it has been extended to black holes in the presence of dark energy fields.

Dark Matter (DM) and Dark Energy (DE) are two prominent issues in modern cosmology and astrophysics. Investigating these components facilitates understanding of the structure and evolution of the universe. Astronomical observations indicate that the universe is expanding at an accelerated rate due to DE, which involves a negative pressure driving the expansion. The introduction of DE has provided a reasonable explanation for this accelerated expansion~\cite{Q1,Q2,Q3}. Statistically, DE accounts for about 73\% of the total energy content of the universe, DM for about 23\%, and ordinary baryonic matter only about 4\%. Despite extensive astronomical observations, the fundamental nature and origin of DE remain a mystery. Various models have been proposed to explain DE, such as quintessence and the cosmological constant. Quintessence is characterized by an equation relating pressure $P$ and energy density $\rho_q$, expressed as $P = \omega_q \rho_q$. Depending on the value of the quintessence state parameter $\omega_q$, three different scenarios arise: phantom energy ($\omega_q < -1$), the cosmological constant ($\omega_q = -1$), and quintessence-like matter ($-1 < \omega_q < -1/3$).

Black hole metrics surrounded by quintessence have long been a popular topic. Kiselev first proposed a Schwarzschild black hole solution in the presence of a quintessence field in~\cite{Q4}. Following this work, many black hole solutions surrounded by typical dark energy fields have been derived. Toshmatov et al. extended Kiselev’s solution to rotating black holes using the Newman-Janis algorithm~\cite{Q5,Q6}, as shown in~\cite{Q7}. Xu et al. also applied similar methods to obtain the corresponding Kerr–Newman–AdS solution~\cite{Q8}. By examining the thermodynamic consistency of these black holes, Ref. \cite{Q9} refined and further generalized these solutions. In \cite{Q10}, Benavides-Gallego et al. considered a rotating black hole with magnetic charge surrounded by quintessence and further analyzed the event horizon, ergosphere, and angular velocity of zero angular momentum observers. More importantly, they argued that dark energy significantly affects black hole dynamics. Moreover, Ndogmo et al. investigated the thermodynamic behavior of rotating black holes with nonlinear magnetic charge in a quintessence field, confirming the second law of thermodynamics and revealing phase transitions~\cite{Q20}.

The structure of this paper is organized as follows: in the next section, we investigate the topological class of Schwarzschild black hole surrounded by quintessence. Section III shifts the focus to the thermodynamic topological classification of Kerr black hole in the presence of quintessence. Sections IV and V examine the Schwarzschild–AdS and Kerr–AdS black holes within the quintessence matter, respectively. Section VI discusses the topological characteristics of Einstein–Gauss–Bonnet (EGB) black holes. The final section provides our conclusions.
\section{Schwarzschild black hole surrounded by quintessence}
We begin by reviewing the work of Kiselev ~\cite{Q4}, who obtained an exact solution to the Einstein field equations representing a static and spherically symmetric black hole surrounded by quintessence matter. In his formulation, the spacetime metric expresses the general form:
\begin{equation}
ds^2=-f(r)dt^2+\frac{1}{f(r)}dr^2+r^2d\Omega _{2}^{2},
\end{equation}
for Schwarzschild black hole, $N(r)$ can be infored as
\begin{equation}
f(r)=1-\frac{2M}{r} -\frac{c}{r^{3\omega _q+1}},
\end{equation}
here, $M$ denotes the black hole mass, and $c$ is a positive normalization factor linked to the quintessence energy density. When the equation-of-state parameter lies within the interval $-1<\omega _q <-1/3$, quintessence effectively accounts for the observed cosmic acceleration. Note that when $c=0$ the metric reduces to the standard Schwarzschild black hole metric. In this paper, we focus on a typical dark energy model and adopt the representative parameter $\omega _q=-2/3$. For this quintessence model, the energy-momentum tensor takes the following form
\begin{equation}
\begin{aligned}
T_t^t & =T_r^r=-\rho, \\
T_\theta^\theta & =T_\phi^\phi=\frac{\rho}{2}\left(3 \omega_q+1\right),
\end{aligned}
\end{equation}
density $\rho_q$ is written as
\begin{equation}
\begin{aligned}
\rho_q  =-\frac{3}{2} \frac{c \omega_q}{r^{3\left(\omega_q+1\right)}} .
\end{aligned}
\end{equation}
The thermodynamic quantities related to the mass, Hawking temperature and entropy of this black hole are
\begin{equation}\label{M}
M=\frac{r_+}{2} -\frac{c}{2} r_+^{-3\omega_q },
\end{equation}
\begin{equation}
T=\frac{1}{4\pi r_+} (1+\frac{3c \omega_q }{r_{+}^{3 \omega_q+1} } ),
\end{equation}
\begin{equation}\label{S}
S=\pi r_+^2.
\end{equation}
It is straightforward to confirm that, for the Schwarzschild black hole surrounded by quintessence (Sch–BH-$\omega_q$), the Hawking temperature tends to vanish as the event horizon radius $r_{+}$ approaches infinity, whereas it becomes divergent when $r_{+}$ approaches the minimum horizon radius $r_{m}$. As a result, the inverse temperature $\beta(r_{+})$ shows
 \begin{equation}
\beta (r_{m})=0, \beta (r_{\infty})=\infty.
\end{equation}
At above asymptotic limits, defect curve $\beta (r_{+})$ is analytical in the range $(r_{m}, \infty)$.
The topological properties of the Sch–BH-$\omega_q$ are fundamentally rooted in the concept of generalized off shell free energy. According to eqs. (5-7), the form of the free energy is written as~\cite{Wei2}
\begin{equation}
\begin{aligned}
\mathcal{F} &=M-\frac{S}{\tau}=\frac{1}{2}(r_+-cr_+^{-3\omega_q })-\frac{\pi r_+^2}{ \tau },
\end{aligned}
\end{equation}
$\tau$ denotes an auxiliary parameter that can be interpreted as the inverse temperature parameter of the cavity surrounding the black hole. The free energy corresponds to its on-shell value when $\tau=\beta=1/T$. According to Duan's $\phi$-mapping theory \cite{Duan1,Duan2}, we introduce an additional parameter $\Theta\in (0, \pi)$, accordingly, the two-component vector $\phi$ is specified by the gradient of as $\tilde{\mathcal{F}}=\mathcal{F}+\frac{1}{\sin \Theta}$ ~\cite{T14}
\begin{equation}
\phi=(\phi ^{r_+}, \phi ^\Theta)=(\partial _{r_+}\tilde{\mathcal{F}}, \partial _{\Theta}\tilde{\mathcal{F}}),
\end{equation}
$\Theta$ is an auxiliary factor introduced for convenience in topological calculations. The value $\Theta=\pi/2$ orresponds to the zero point of the vector field $\phi$. Based on $\phi$-mapping theory, each black hole state is assigned a winding number reflecting its topological charge. Locally stable states correspond to $w=+1$, while unstable ones have $w=-1$. The sum of all local winding numbers defines the global topological number $W$, which determines the black hole’s overall topological class.

We establish the components of the vector field $\phi$ as:
\begin{equation}
\phi ^{r_+}=\frac{1}{2}(1+3c \omega _q r_+^{-1-3\omega _q}) -\frac{2\pi r_+}{\tau },
\end{equation}
\begin{equation}
\phi ^\Theta =-\mathrm { cot} \Theta \mathrm {csc} \Theta.
\end{equation}
The inverse temperature parameter can be written by using the equation $\phi ^{r_h}=0$
\begin{equation}\label{tt}
  \tau =\beta=\frac{4\pi r_+^{2+3\omega _q}}{r_+^{1+3\omega _q}+3c\omega _q}.
\end{equation}
\begin{figure}[]
		\centering
		\includegraphics[width=2.8in,height=2.1in]{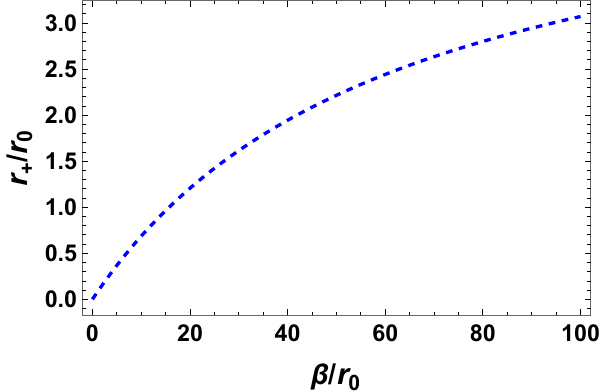}
     \caption{The zero points of $\phi ^{r_+}$ in the $r_+-\beta$ plane for Sch-BH-$\omega_q$, we set $a/r_0=0.1$, $c r_0^2=0.1$.}
\end{figure}
Figure 1 presents the component $\phi ^{r_+}$ in the $r_+-\beta$ plane, here $r_0$ denotes an arbitrary length scale determined by the size of the cavity enclosing the black hole. The figure shows that the Sch–BH-$\omega_q$ possesses a single unstable branch. We proceed to examine the variation of the vector field $\phi$ (see Fig. 2) along the boundary defined by Eq. (8). The complete boundary is outlined by the closed contour $C=I_{1}\cup I_{2}\cup I_{3}\cup I_{4}$, whose segments and overall shape are shown in Fig. 3, encompassing the entire parameter domain. Due to the construction of $\phi$, it remains perpendicular to $I_{2}$ and $I_{4}$~\cite{T14}, implying that the key asymptotic properties are determined by its behavior on $I_{1}$ and $I_{3}$. Specifically, as $r_{+} \to r_{m}$, the vector field directs rightward with an inclination governed by $\phi ^\Theta$; whereas in the limit $r_{+} \to \infty $, it turns leftward, as illustrated in Fig. 2 and Fig. 3. Table I presents the orientations of the vector field $\phi$ along segments $I_{1}$ and $I_{3}$.

The topological charge of the Sch–BH–$\omega_q$ can be determined by examining the behavior of the vector field $\phi = (\phi^{r_+}, \phi^{\Theta})$ along the contour displayed in Fig. 2. As depicted in Fig. 4, from the variations of $\phi$ along contour $C_1$, one can observe that the resulting closed curve $\Phi_1$ rotates clockwise, indicating that the zero point (marked by the black dot) possesses a winding number of -1. The corresponding global topological charge $W$ is listed in Table I.

\begin{figure}[H]
		\centering
		\includegraphics[width=2.8in,height=2.5in]{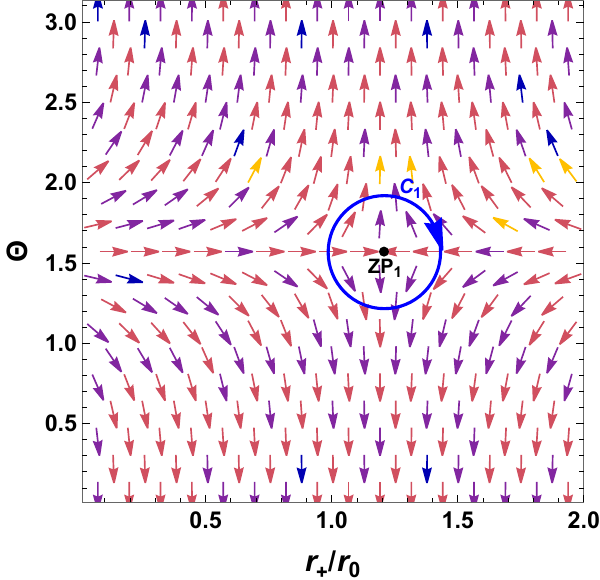}
     \caption{The unit vector field $n$ on the $r_+-\Theta$ plane for the Sch-BH-$\omega_q$, we set $\beta/r_0=20$.}
\end{figure}

\begin{figure}[htbp]
		\centering
		\includegraphics[width=2.8in,height=2.1in]{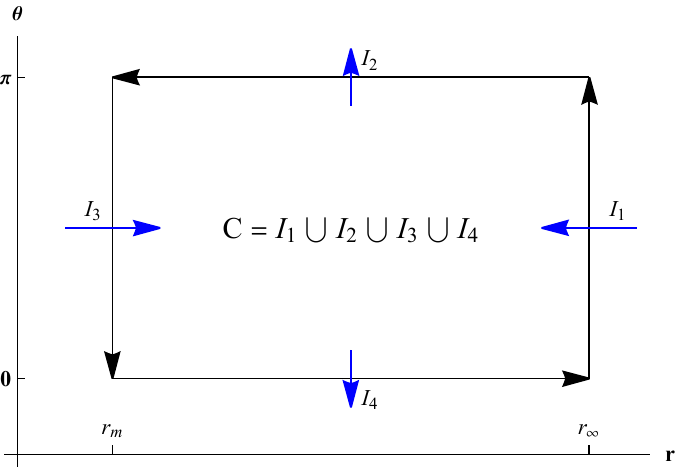}
     \caption{The asymptotic behavior of the $n$-vector field $\phi$ at the boundary ($C=I_{1}\cup I_{2}\cup I_{3}\cup I_{4}$).}
\end{figure}

\begin{figure}[htbp]
		\centering
		\includegraphics[width=2.8in,height=2.1in]{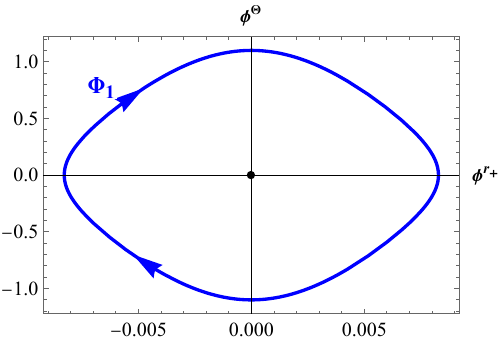}
     \caption{Contour $\Phi_{1}$ depicts the variation of the vector field $\phi$ components along the pat $C_1$ show in Fig. 2 for the Sch-BH-$\omega_q$ case.}
\end{figure}

\begin{table}[]
	\centering
\caption{The orientation is indicated by the arrow of $\phi^{r_+}$ for the Sch-BH-$\omega_q$ and Kerr-BH-$\omega_q$ and the topological numbers.}
\begin{tabular}{c|p{1cm}<{\centering}|p{1cm}<{\centering}|p{1cm}<{\centering}|p{1cm}<{\centering}|p{1cm}<{\centering}}
 \hline \hline BH spacetime                          &$I_{1}$                         & $I_{2}$                 & $I_{3}$ & $I_{4}$ & $W$\\
 \hline
       Sch-BH-$\omega_q$ &$\gets$ &$\uparrow$  &$\to$   &$\downarrow$     &-1 \\
       Kerr-BH-$\omega_q$           &$\gets$ &$\uparrow$  &$\gets$ &$\downarrow$     &0   \\
\hline \hline
\end{tabular}
\end{table}
The universal thermodynamic properties of the Sch-BH-$\omega_q$ can be described as follows. In the low-temperature limit ($\beta \to \infty$), large black holes appear but remain thermodynamically unstable because of their negative topological number. In contrast, at high temperatures ($\beta \to 0$), small black holes emerge, which are also unstable. Hence, according to the general thermodynamic classification framework proposed in Ref.~\cite{T14}, the Sch–BH–$\omega_q$ belongs to the $W^{1-}$ category.
\section{Kerr black hole surrounded by quintessence}
This section investigates the topological classification of Kerr spacetime under a quintessence field. First, in the Boyer-Lindquist coordinate system, the Kerr spacetime metric in the quintessence background is given by:~\cite{kbh}
\begin{equation}\label{kbh}
\begin{aligned}
d s^2= & -\left(1-\frac{2 M r+c r^{1-3 \omega_q}}{\Sigma}\right) d t^2+\frac{\Sigma}{\Delta} d r^2\\
&-2 a \sin ^2 \theta\left(\frac{2 M r+c r^{1-3 \omega_q}}{\Sigma}\right) d \phi d t+\Sigma d \theta^2 \\
& +\sin ^2 \theta\left[r^2+a^2+a^2 \sin ^2 \theta\left(\frac{2 M r+c r^{1-3 \omega_q}}{\Sigma}\right)\right] d \phi^2,
\end{aligned}
\end{equation}
\begin{equation}
\begin{aligned}
&\Delta=r^2-2 M r+a^2-c r^{1-3 \omega_q}, \Sigma=r^2+a^2cos^2\theta.
\end{aligned}
\end{equation}
Where $a$ is the rotation parameter, in the absent of the quintessence field $c$, the spacetime metric \eqref{kbh} coincides with Kerr metric. The mass, temperature and entropy of a black hole are respectively functions of the event horizon radius $r_+$
\begin{equation}
M=\frac{1}{2} (r_+-cr_+^{-3\omega _q}+\frac{a^2}{r_+}),
\end{equation}
\begin{equation}
T=\frac{r_+^2-a^2+3c\omega _qr_+^{1-3\omega _q}}{4\pi r_+(r_+^2+a^2)},
\end{equation}
\begin{equation}
S=\pi(r_+^2+a^2),
\end{equation}
the asymptotic behavior of the inverse temperature $\beta(r_{+})$ for the Kerr black hole surrounded by quintessence (Kerr-BH-$\omega_q$) has the limits
\begin{equation}
\beta (r_{m})=\infty, \beta (r_{\infty})=\infty.
\end{equation}

Based on above thermodynamic quantities, the expression for the free energy can be derived
\begin{equation}
\mathcal{F} =\frac{1}{2} (r_+-cr_+^{-3\omega _q}+\frac{a^2}{r_+})-\frac{\pi(r_+^2+a^2)}{\tau }.
\end{equation}
The components of the vector field $\phi$ can be calculated as follows
\begin{equation}
\begin{aligned}
\phi^{r_+}&=\frac{1}{2} (1-\frac{a^2}{r_+^2} +3c\omega _qr_+^{-1-3\omega _q})-\frac{2\pi r_+}{\tau },\\
\phi ^\theta & =-\mathrm { cot} \Theta \mathrm {csc} \Theta.
\end{aligned}
\end{equation}
When the vector component $\phi^{r_+}$ is zero, we can obtain
\begin{equation}
\tau =\beta=\frac{4\pi r_+^{3+3\omega _q}}{r_+^{2+3\omega _q}-a^2r_+^{3\omega  _q}+3c\omega _qr_+}.
\end{equation}
\begin{figure}[]
		\centering
		\includegraphics[width=2.8in,height=2.1in]{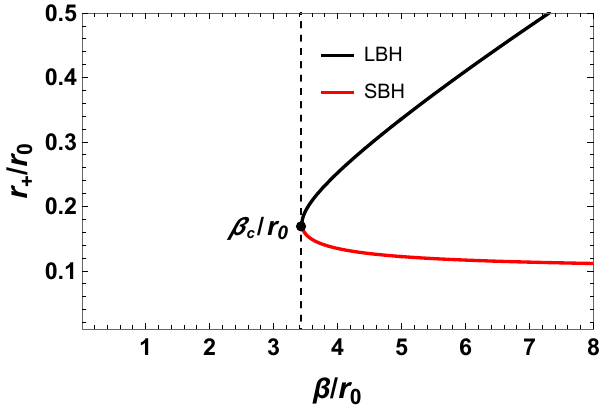}
     \caption{The zero points of $\phi ^{r_+}$ in the $r_+-\beta$ plane for Kerr-BH-$\omega_q$, we set $a/r_0=0.1$, $c r_0^2=0.1$. }
\end{figure}

\begin{figure}[htbp]
		\centering
		\includegraphics[width=2.8in,height=2.5in]{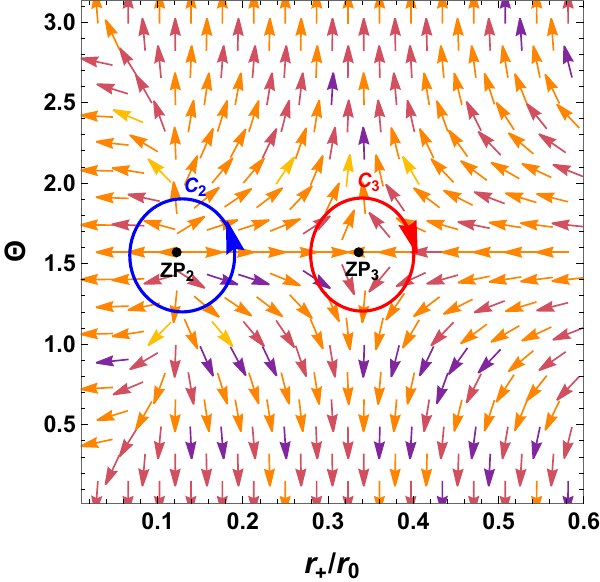}
     \caption{The unit vector field $n$ on the $r_+-\Theta$ plane for the Kerr-BH-$\omega_q$ for $\beta/r_0=5$.}
\end{figure}

\begin{figure}[htbp]
		\centering
		\includegraphics[width=2.8in,height=2.1in]{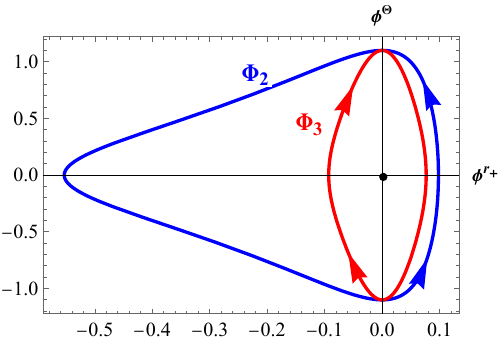}
     \caption{Contours $\Phi_{2}$ and $\Phi_{3}$ depict the variation of the vector field $\phi$ components along the pats $C_2$ and $C_3$ show in Fig. 6 for the Kerr-BH-$\omega_q$ case.}
\end{figure}
The plot of $r_+$ versus $\beta$ is shown in Fig. 5, where a small black hole branch and a large black hole branch can be observed. A special point $\beta_c=\frac{12\sqrt{3}\pi a}{2-6\sqrt{3}ca }$ for which $d\beta/dr_+=0$ is degenerate, at this point we find that $(d^2\beta/dr_+^2)>0$, indicating that $\beta_a$ is a generation point. When we set $a/r_0=0.1$, $c r_0^2=0.1$, the generation point located in $\beta/r_0=\beta_c/r_0=3.4438$.

The characteristics of the vector field $n$ shown in Fig. 6, together with its asymptotic behavior along the boundary segments $I_i$ (analogous to the case of the Sch-BH-$\omega_q$ illustrated in Fig. 3), are summarized in Table I. It is found that in the limits $r_h \to r_m$ or $r_h \to \infty$, the vector field plot of $\phi$ points leftward.

Figure 7 illustrates the variation pattern of the vector field $\phi$ along the chosen contours. It can be observed that the closed trajectory $\Phi_2$ rotates counterclockwise, leading to winding numbers of +1 and -1 for the first and second zero points, respectively. This result indcates a total topological number $W=0$ for the Kerr-BH-$\omega_q$.

For the universal thermodynamic behavior, when $\beta \to \infty$, both a stable small black hole and an unstable large black hole coexist. In contrast, as $\beta \to 0$, the black hole phase disappears entirely. Therefore, the Kerr-BH-$\omega_q$ is identified with the topological class $W^{0+}$. Moreover, the winding numbers and universal topological class for the Kerr-BH-$\omega_q$ match those of the standard Kerr case \cite{T17}, yet they deviate from the values of the Schwarzschild black hole under the same dark energy influence. Therefore, the spacetime rotation in the quintessence framework influences its topological characteristics.

\section{Schwarzschild-adS black hole surrounded by quintessence}
Then using the ideas of Kiselev who proposed the solution of the Schwarzschild–anti de Sitter black hole surrounded by quintessence (Sch-aBH-$\omega_q$) ~\cite{Q4}
\begin{equation}
ds^2=-f(r)dt^2+\frac{dr^2}{f(r)}+r^2d\theta ^{2}+r^2\sin^2\theta d \phi^2,
\end{equation}
with
\begin{equation}
f(r)=1-\frac{2M}{r} -\frac{\Lambda  r^2}{3}-\frac{c}{r^{3\omega _q+1}},
\end{equation}
where $\Lambda$ is the cosmological constant. Then by using Eq. (24), one gets
\begin{equation}
M=\frac{r_+}{2} -\frac{cr_+^{-3\omega_q }}{2} -\frac{\Lambda r_+^3}{6}.
\end{equation}
The Hawking temperature and entropy associated with the black hole horizon can be determined
\begin{equation}
\begin{aligned}
&T=\frac{1}{4\pi r_+} -\frac{\Lambda r_+}{4\pi} +\frac{3c\omega_q r_+^{-2-3\omega_q }}{4\pi},\\
&S=\pi r_+^2.
\end{aligned}
\end{equation}
The asymptotic behavior of the inverse temperature $\beta (r_{+})$ exhibits
\begin{equation}
\beta (r_{m})=0,~\beta (r_{\infty})=0.
\end{equation}
The generalized off-shell Helmholtz free energy can be easily obtained
\begin{equation}
\mathcal{F}=\frac{r_+}{2} -\frac{cr_+^{-3\omega_q }}{2} -\frac{\Lambda r_+^3}{6} -\frac{\pi r_+^2}{\tau },
\end{equation}
the components of the vector $\phi$ can be computed as
\begin{equation}
\begin{aligned}
\phi^{r_+}=\frac{1}{2} +4\pi P r_+^2+\frac{3c\omega_q r_+^{-1-3\omega_q }}{2} -\frac{2\pi r_+}{\tau },
\end{aligned}
\end{equation}
\begin{equation}
\phi ^\theta =-\cot \Theta \csc \Theta,
\end{equation}
and the inverse temperature is
\begin{equation}
\beta=\frac{4\pi r_+^{2+3\omega_q}}{r_+^{1+3\omega_q}+8\pi P r_+^{3+3\omega_q}+3c\omega_q}.
\end{equation}
\begin{figure}[]
		\centering
		\includegraphics[width=2.8in,height=2.1in]{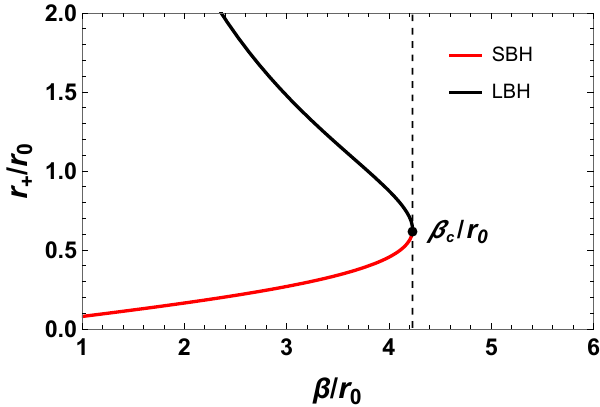}
     \caption{The zero points of $\phi ^{r_+}$ in the $r_+-\beta$ plane for the Sch-aBH-$\omega_q$, we set $a/r_0=0.1$, $c r_0^2=0.1$, and $Pr_0^2=0.1$.}
\end{figure}
\begin{figure}[htbp]
		\centering
		\includegraphics[width=2.8in,height=2.5in]{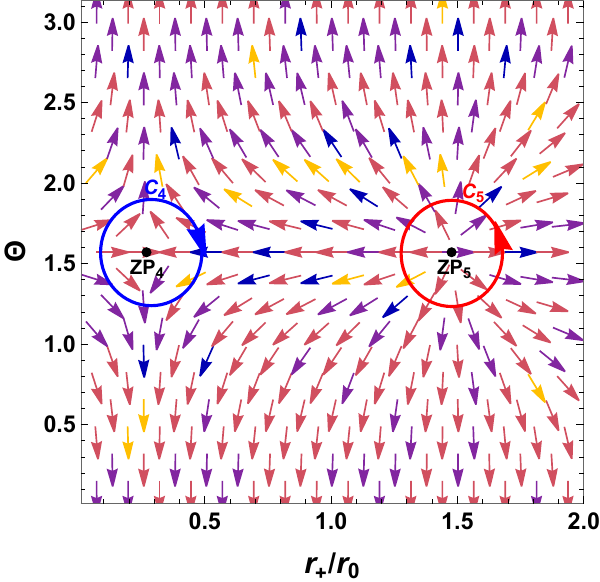}
     \caption{The unit vector field $n$ on the $r_+-\Theta$ plane for the Sch-aBH-$\omega_q$, we set $\beta/r_0=3$.}
\end{figure}

\begin{figure}[htbp]
		\centering
		\includegraphics[width=2.8in,height=2.1in]{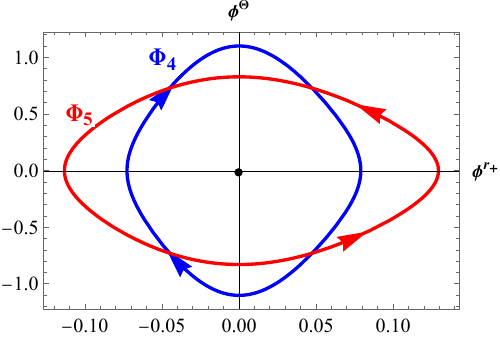}
     \caption{Contours $\Phi_{4}$ and $\Phi_{5}$ depict the variation of the vector field $\phi$ components along the pats $C_4$ and $C_5$ show in Fig. 9 for the Sch-aBH-$\omega_q$ case.}
\end{figure}
In this framework, the pressure term is defined by $P = -\Lambda / 8\pi$. The plot of $r_+$ versus $\beta$ is shown in Fig. 8. At the point $\beta_c=\frac{2\pi }{3\sqrt{2\pi P} -c}$ which is a degenerate point, we find that $(d^2\beta/dr_+^2)<0$, indicating that $\beta_c$ is an annihilation point. When $\beta < \beta_c$, two black hole phases coexist a small branch and a large branch. Once $\beta > \beta_c$, these phases disappear, leaving no viable black hole solution. The annihilation point locates at $\beta_c/r_0=4.23016$ for $c r_0^2=0.1$ and $Pr_0^2=0.1$.

The corresponding vector field $n$ associated with the Sch-aBH-$\omega_q$ is illustrated in Fig. 9. As shown in the figure, this black hole exhibits two distinct zero points (depicted as black dots), which is in contrast to the Sch-BH-$\omega_q$ that possesses only one. The asymptotic orientation of the vector field $\phi$ at the boundaries $I_i$, as listed in Table II, reveals that as $r_+ \to r_m$ or $r \to \infty$, the field tends to point rightward. This directional behavior indicates that the thermodynamic topology of the system corresponds to a particular universal class.

To further elucidate the topological structure, Fig. 10 presents the evolution of the vector field components ($\phi^{r_+}, \phi^{\Theta}$) traced along the closed contours $C_i$ depicted in Fig. 9. The zero point of $\Phi_i$ is located at the origin. Contours with the first and second zeros possess winding numbers -1 and +1, respectively.

In summary, the thermodynamic evolution proceeds as follows: at very low temperatures ($\beta \to \infty$), the system does not support any black hole configuration, whereas in the high-temperature limit ($\beta \to 0$), two distinct phases appear an unstable small black hole and a stable large black hole. Therefore, the Sch–aBH–$\omega_q$ is identified as belonging to the $W^{0-}$ topological category.
\begin{table}[]
	\centering
\caption{The orientation is indicated by the arrow of $\phi^{r_+}$ for the Sch-aBH-$\omega_q$ and Kerr-aBH-$\omega_q$ and the topological numbers..}
\begin{tabular}{c|p{1cm}<{\centering}|p{1cm}<{\centering}|p{1cm}<{\centering}|p{1cm}<{\centering}|p{1cm}<{\centering}}
 \hline \hline BH spacetime                          &$I_{1}$                         & $I_{2}$                 & $I_{3}$ & $I_{4}$ & $W$\\
 \hline
       Sch-aBH-$\omega_q$ &$\to$  &$\uparrow$  &$\to$   &$\downarrow$     &0 \\
      Kerr-aBH-$\omega_q$        &$\to$  &$\uparrow$  &$\gets$ &$\downarrow$     &1  \\
\hline \hline
\end{tabular}
\end{table}

\section{Kerr-adS black hole surrounded by quintessence}
Next, we turn to the topological class of the Kerr-anti de Sitter black hole surrounded by quintessence (Kerr-aBH-$\omega_q$). In Boyer-Lindquist coordinates, the spacetime of this black hole is given by~\cite{kabh}
\begin{equation}
\begin{aligned}
d s^2= & -\frac{\Delta_r}{\Omega}\left(d t-\frac{a \sin ^2 \theta}{\Xi} d \phi\right)^2+\frac{\Omega}{\Delta_r} d r^2 \\
& +\frac{\Omega}{\breve{P}} d \theta^2+\frac{\breve{P} \sin ^2 \theta}{\Omega}\left(a d t-\frac{\left(r^2+a^2\right)}{\Xi} d \phi\right)^2,
\end{aligned}
\end{equation}
where the relevant terms are defined as
\begin{equation}
\begin{aligned}
\Delta_r=r^2+a^2+\left(r^2+a^2\right)\left(\frac{r^2}{l^2}\right)-2 m r-c r^{1-3 w},
\end{aligned}
\end{equation}
\begin{equation}
\Omega =r^2+a^2\cos^2\theta ,\quad \Xi =1-\frac{a^2}{l^2} ,\quad \breve{P}=1-\frac{a^2}{l^2}\cos^2\theta.
\end{equation}
Under the horizon condition $\Delta_r=0$, the black hole mass, entropy, and Hawking temperature are
\begin{equation}
M=\frac{m}{\Xi^2},
\end{equation}
\begin{equation}
S=\frac{\pi(a^2+r_+^2)}{\Xi },
\end{equation}
\begin{equation}
T=\frac{\Delta r^\prime (r)}{4\pi(r^2+a^2)}\bigg|_{{r=r_+ }}.
\end{equation}
Using the relation $P=\frac{3}{8\pi l^2}$, the generalized free energy takes the form
\begin{equation}
\begin{aligned}
\mathcal{F} =&\frac{3(3+8\pi Pr_+^2)(r_+^2+a^2)-9cr_+^{1-3\omega q}}{6r_{+}(3-8\pi P a^2)^2}\\
&-\frac{3\pi (r_+^2+a^2)}{(3-8\pi P a^2)\tau }.
\end{aligned}
\end{equation}
and the components of the vector $\phi$  can be yielded as:
\begin{equation}
\begin{aligned}
\phi^{r_+}=&\frac{9(r_+^2-a^2)+24\pi Pr_+^2(3r_+^2+a^2)+27c\omega_qr_{+}^{1-3\omega_q}}{2r_{+}^2(3-8\pi Pa^2)^2}\\
&-\frac{6 \pi r_+}{3-8\pi P a^2\tau },
\end{aligned}
\end{equation}
\begin{equation}
\phi ^\theta =-\cot \Theta \csc \Theta,
\end{equation}
with
\begin{widetext}
\begin{equation}
\beta =\frac{4\pi r_+^{3+3\omega_q}(3-8\pi Pa^2)}   {3r_+^{2+3\omega_q}+24\pi Pr_+^{4+3\omega_ q}-3a^2r_+^{3\omega_q}+8\pi Pa^2r_+^{2+3\omega _q}+9c\omega_qr_+}.
\end{equation}
\end{widetext}
Figure 11 shows the variation of $r_+$ with $\beta$, a generation point and an annihilation point located in $\beta/r_0=\beta_a/r_0=3.0101$ and $\beta/r_0=\beta_b/r_0=4.2354$, respectively. These points divide the black hole into three branches: a small black hole branch, an intermediate black hole branch, and a large black hole branch.
\begin{figure}[]
		\centering
		\includegraphics[width=2.8in,height=2.1in]{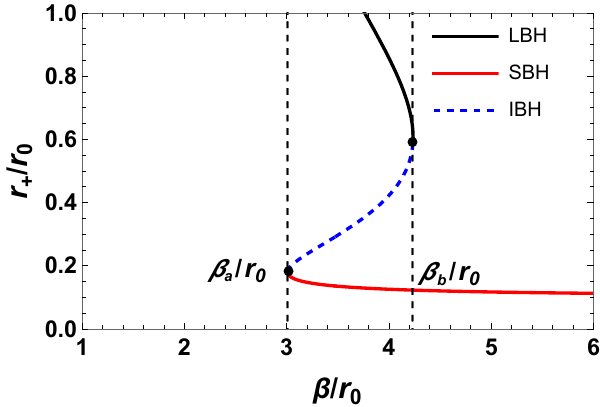}
     \caption{The zero points of $\phi ^{r_+}$ in the $r_+-\beta$ plane for the Kerr-aBH-$\omega_q$, we set $a/r_0=0.1$, $c r_0^2=0.1$ and $P r_0^2=0.1$.}
\end{figure}
The characteristics of the vector field $n$ shown in Fig. 12, where three zero points, indicated by black dots, are identified. These zeros partition the black hole configurations into small, intermediate, and large types. Fig. 13 presents the evolution of the vector field components ($\phi^{r_+}, \phi^{\Theta}$) traced along the closed contours $C_i$ depicted in Fig. 12. The zero point of $\Phi_i$ is located at the origin. Contours with three zeros (from left to right) possess winding numbers +1, -1 and +1, respectively. The asymptotic behavior of $\phi$ along the boundary segments $I_i$ is summarized in Table II: at $I_1$ the vector field points rightward, while at $I_3$ it points leftward, revealing its topological number of +1. In the regime of low temperature ($\beta \to \infty$), the configuration gives rise to stable small black holes; conversely, when $\beta \to 0$, stable large black holes dominate the system. Thus, the Kerr-aBH-$\omega_q$ is classified within the topological class $W^{1+}$.

\begin{figure}[htbp]
		\centering
		\includegraphics[width=2.8in,height=2.5in]{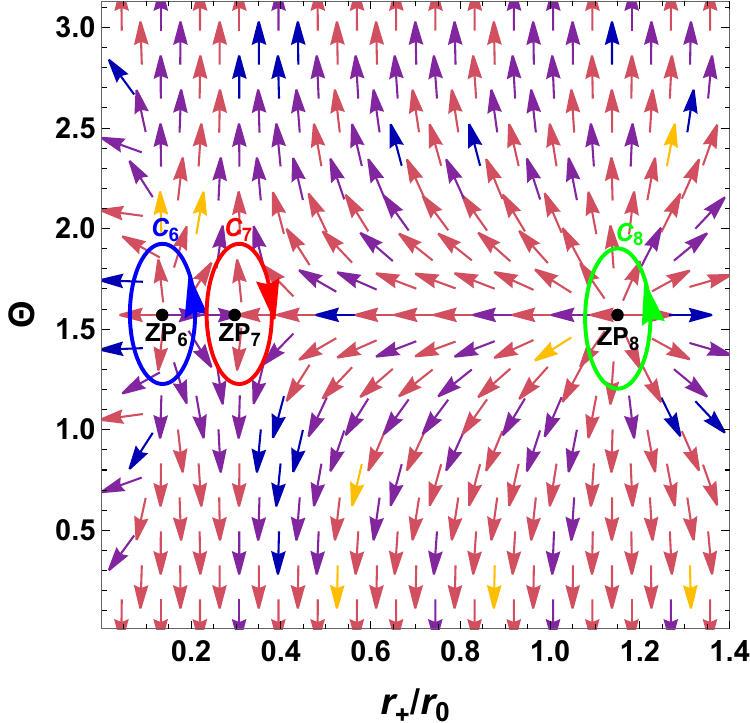}
     \caption{The unit vector field $n$ on the $r_+-\Theta$ plane for the Kerr-aBH-$\omega_q$, we set $\beta/r_0=3.5$.}
\end{figure}
\begin{figure}[]
		\centering
		\includegraphics[width=2.8in,height=2.1in]{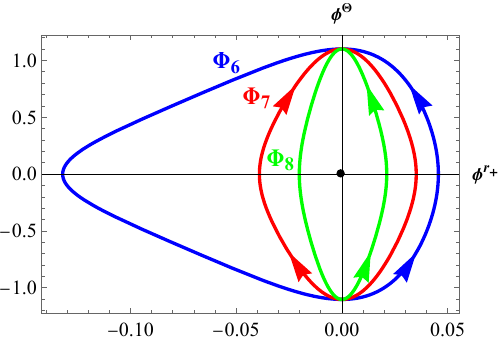}
     \caption{Contours $\Phi_{6}$, $\Phi_{7}$ and $\Phi_{8}$ depict the variation of the vector field $\phi$ components along the pats $C_6$, $C_7$ and $C_8$ show in Fig. 12 for the Kerr-aBH-$\omega_q$.}
\end{figure}

\section{Einstein-Gauss-Bonnet black hole surrounded by quintessence}
In this section, we focus on the thermodynamic topological number associated with black hole entropy in complex forms, such as black hole solutions surrounded by dark energy in the Einstein-Gauss-Bonnet (EGB) gravity theory. The EGB action is given by:
\begin{equation}
S=\frac{1}{16 \pi} \int d^D x \sqrt{-g}\left[R+\alpha L_{G B}\right],
\end{equation}
here, $R$ the Richie tensor, $\alpha$ is the Gauss-Bonnet constant, Gauss-Bonnet Lagrange $L_{G B}$ by
\begin{equation}
L_{G B}=R^2-4 R_{a b} R^{a b}+R_{a b c d} R^{a b c d} .
\end{equation}
Consider the four-dimensional EGB black hole solution in the presence of quintessence matter. The corresponding spacetime metric is given by:~\cite{4DQ}
\begin{equation}
d s^2=-f(r) d t^2+\frac{d r^2}{f(r)}+r^2\left(d \theta^2+\sin ^2 \theta d \varphi^2\right),
\end{equation}
where
\begin{equation}\label{EGB1}
\begin{aligned}
f_{\pm} =1+\frac{r^2}{2\alpha }(1\pm \sqrt{1+\frac{8\alpha M}{r^3} +\frac{8\alpha q}{r^{3\omega _q+3}} } ).
\end{aligned}
\end{equation}

It is evident that when $q=0$, the solution simplifies to the four-dimensional EGB black hole proposed by Glavan and Lin ~\cite{Glavan}. In the limit $\alpha\to 0$, the negative branch recovers the Schwarzschild black hole immersed in quintessence matter, as described in ~\cite{Q4}. Taking both $q=0$ and $\alpha\to 0$, the metric reduces to the classical Schwarzschild solution. By adjusting the state parameter $\omega _q$, a variety of black hole configurations can be realized. For example, when $\omega _q=0$, corresponding to the absence of quintessence, the metric again reduces to the standard 4D EGB black hole ~\cite{Glavan}. The function $f(r)$ and the location of the event horizon are highly sensitive to the parameters $\omega _q$, $q$, and the Gauss–Bonnet coupling $\alpha$. To ensure the physical validity of the quintessence field, $\omega _q$ must remain within the interval  $-1<\omega _q<-1/3$, , which ensures a positive energy density of the dark energy component, given by ~\cite{4D}
\begin{equation}
\rho_q=-6 w_q q \frac{1}{r^{3\left(1+w_q\right)}}.
\end{equation}
In equation \eqref{EGB1}, the ``$\pm$'' symbol represents the two branches of the solution. For large horizon radius $r$, equation \eqref{EGB1} reduces to the four-dimensional Schwarzschild black hole solution surrounded by quintessence matter. Therefore, the negative branch of the solution is well-behaved. Consequently, we consider the negative branch solution in the following analysis, from which several thermodynamic quantities of the black hole can be derived as:
\begin{equation}
 M=\frac{r_+}{2} (1+\frac{\alpha }{r_+}-\frac{2q}{r_+^{3\omega _q+1}}  ),
\end{equation}
\begin{equation}
T=\frac{r_{+}^{3 \omega _q+1}\left(r_{+}^2-\alpha\right)+6 \omega _q q r_{+}^2}{4 \pi r_{+}^{3 \omega _q+2}\left(r_{+}^2+2 \alpha\right)}.
\end{equation}
Now, using the first law of thermodynamics $dM=TdS$, we obtain the expression for the black hole entropy
\begin{equation}
S=\frac{A}{4} [1+\frac{4}{r_+^2}\mathrm {log}(r_+)    ],
\end{equation}
here $A=4\pi r_+^2$ represents area of the black hole horizon. Based on above thermodynamic quantities, the generalized off-shell free energy of the system can be derived
\begin{equation}
\mathcal{F}=\frac{r_+}{2} (1+\frac{\alpha }{r_+}-\frac{2q}{r_+^{3\omega _q+1}}  )-\frac{\pi r_+^2( r_+^2+4\alpha \mathrm {log}(r_+)   )}{r_+^2\tau }.
\end{equation}
From this, the component $\phi^{r_+}$ of the vector field $\phi$ can be determined
\begin{equation}
\phi^{r_+}=\frac{1}{2} +3q\omega _qr_+^{-1-3\omega _q}-\frac{2\pi (r_+^2+2\alpha )}{r_+\tau }.
\end{equation}
Therefore, the inverse temperature of the 4D EGB black hole is given by
\begin{equation}
\beta=\frac{4\pi r_+^{3\omega _q}(r_+^2+2\alpha )}{r_+^{1+3\omega _q}+6q\omega _q}.
\end{equation}
Figure 14 presents the zero-point diagram of $\phi^{r_+}$, from which it can be seen that the 4D EGB black hole surrounded by quintessence (4D-EGB-BH-$\omega_q$) features a generation point $\beta_c$ but lacks an annihilation point. Figures 15 and 16 indicate that for $\beta>\beta_c$, the winding numbers of the zero points enclosed by the contours $C_9$ and $C_{10}$ are 1 and -1, respectively, yielding a total topological number $W=0$. This outcome differs from that of the Schwarzschild black hole surrounded by quintessence matter. Nevertheless, it can be inferred that the 4D EGB black hole and the Kerr black hole in the same matter field belong to the identical universal thermodynamic class, $W^{0+}$. Furthermore, the asymptotic behavior of the vector field near the boundaries $I_i$ is summarized in Table III.
\begin{figure}[]
		\centering
		\includegraphics[width=2.8in,height=2.1in]{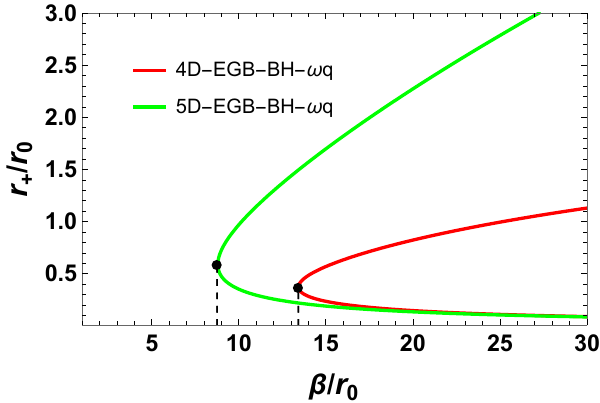}
     \caption{The zero points of $\phi ^{r_+}$ in the $r_+-\beta$ plane for 4D/5D-EGB-BH-$\omega_q$, we set $\alpha/r_0^2=0.1$ and $qr_0=0.1$.}
\end{figure}
\begin{figure}[]
		\centering
		\includegraphics[width=2.8in,height=2.5in]{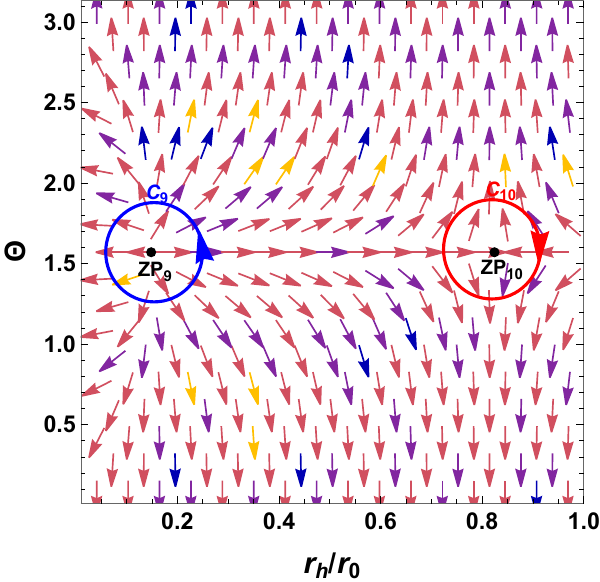}
     \caption{The unit vector field $n$ on the ${r_+}-\Theta$ plane for the 4D-EGB-BH-$\omega_q$, we set $\beta/r_0=20$.}
\end{figure}
\begin{figure}[]
		\centering
		\includegraphics[width=2.8in,height=2.1in]{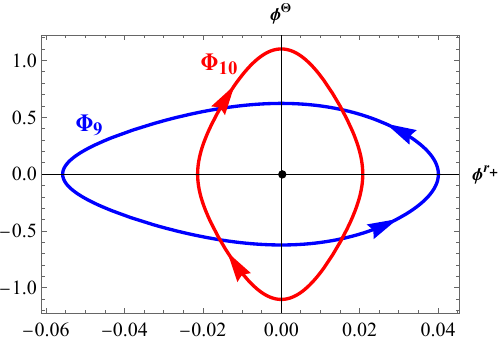}
     \caption{Contours $\Phi_{9}$ and $\Phi_{10}$ depict the variation of the vector field $\phi$ components along the pats $C_9$ and $C_{10}$ show in Fig. 15 for the 4D-EGB-BH-$\omega_q$ case.}
\end{figure}

For the case of 5D EGB black hole surrounded by quintessence (5D-EGB-BH-$\omega_q$), the solution of this black hole is~\cite{5DQ}
\begin{equation}
\begin{aligned}
f_{\pm} =1+\frac{r^2}{4\alpha }(1\pm \sqrt{1+\frac{8\alpha M}{r^4} +\frac{8\alpha q}{r^{4\omega _q+4}} } ),
\end{aligned}
\end{equation}
Likewise, by focusing solely on the negative branch, the thermodynamic quantities associated with this spacetime can be expressed as follows
\begin{equation}
M=r_+^2(1-\frac{q}{r_+^{4\omega _q+2}} +\frac{2\alpha }{r_+^2} ),
\end{equation}
\begin{equation}
S=\frac{4\pi r_+^3}{3} +16\pi \alpha r_+,
\end{equation}
\begin{equation}
T=\frac{1}{2 \pi r_{+}}\left[\frac{1+\frac{2 q \omega_q}{r_{+}^{4 \omega_q+2}}}{1+\frac{4 \alpha}{r_{+}^2}}\right] .
\end{equation}
For this black hole, the off-shell free energy is given by:
\begin{equation}
\mathcal{F}=r_+(1-\frac{q}{r_+^{4\omega _q}} +\frac{2\alpha }{r_+^2} )-\frac{4\pi r_+^3 +48\pi \alpha r_+}{3\tau},
\end{equation}
which in turn yields the following expression for the $\phi^{r_+}$ component of the vector field
\begin{equation}
\phi^{r_+}=2r_++4q\omega _qr_+^{-1-4\omega _q}-\frac{4\pi (r_+^2+4\alpha )}{\tau },
\end{equation}
and
\begin{equation}
\tau=\frac{2\pi r_+^{1+4\omega _q}(r_+^2+4\alpha )}{ r_+^{2+4\omega _q}+2q\omega _q}.
\end{equation}

As shown in Fig. 14, we observe that the zeros of the vector field component $\phi^{r_+}$ for both 4D and 5D EGB black holes exhibit similar behavior. The results for the 5D EGB black hole can be derived from those of the 4D EGB black hole. Therefore, the 5D-EGB-BH-$\omega_q$ belongs to the specific topological class, $W^{0-}$.

\begin{table}[H]
	\centering
\caption{The orientation is indicated by the arrow of $\phi^{r_+}$ for the 4D/5D-EGB-BH-$\omega_q$ and the topological numbers..}
\begin{tabular}{c|p{1cm}<{\centering}|p{1cm}<{\centering}|p{1cm}<{\centering}|p{1cm}<{\centering}|p{1cm}<{\centering}}
 \hline \hline BH spacetime                          &$I_{1}$                         & $I_{2}$                 & $I_{3}$ & $I_{4}$ & $W$\\
 \hline
       4D-EGB-BH-$\omega_q$        &$\gets$ &$\uparrow$  &$\gets$ &$\downarrow$    & 0  \\
       5D-EGB-BH-$\omega_q$         &$\gets$ &$\uparrow$  &$\gets$ &$\downarrow$    & 0   \\
\hline \hline
\end{tabular}
\end{table}
\begin{widetext}
\begin{table}[H]
	\centering
\caption{In the table below, we summarize the universal thermodynamic topological classifications of the black holes surrounded by quintessence, together with a summary of their corresponding thermodynamic behaviors and their generation points (GP) and annihilation points (AP).}
\begin{tabular}{c|p{1.5cm}<{\centering}|p{1.5cm}<{\centering}|p{1.5cm}<{\centering}|p{3cm}<{\centering}|p{3cm}<{\centering}|p{2.5cm}<{\centering}}
 \hline \hline Black hole spacetime             &W Classes   & Innermost   &  Outermos &Low T       & High T        & DP \\
 \hline      Sch-BH \cite{T14}       & $W^{1-}$      &unstable   &  unstable & unstable large      &   unstable small              & in pairs   \\
             Kerr-BH \cite{T17}                  &$W^{0+}$        &stable &unstable&unstable large + stable small & no                  & one more GP  \\
             Sch-aBH \cite{T14}& $W^{0-}$ &   unstable    & stable  &   no                & unstable small + stable large                  &one more AP \\
 Sch-BH-$\omega_q$   & $W^{1-}$ &unstable   &  unstable & unstable large      &   unstable small              & in pairs   \\
 Kerr-BH-$\omega_q$             &$W^{0+}$  & stable   &  unstable   &   stable small + unstable large                & no                  & one more GP   \\
 Sch-aBH-$\omega_q$ & $W^{0-}$ &   unstable    & stable  &   no                & unstable small + stable large                  &one more AP \\
 Kerr-aBH-$\omega_q$          &$W^{1+}$   &stable & stable & stable small         &stable large              & in pair \\
 4D-EGB-BH-$\omega_q$              &$W^{0+}$  &stable &unstable&unstable large + stable small & no                  & one more GP  \\
 5D-EGB-BH-$\omega_q$              &$W^{0+}$   &stable &unstable&unstable large + stable small & no                  & one more GP  \\
\hline \hline
\end{tabular}
\end{table}
\end{widetext}
\section{Conclusions}

In this work, we have systematically investigated the universal thermodynamic topological classifications of a family of black holes surrounded by quintessence, including Schwarzschild, Kerr, AdS, and Einstein–Gauss–Bonnet (EGB) spacetimes. The results are summarized in Table IV. Our analysis reveals that the Schwarzschild black hole in a quintessence background belongs to the $W^{1-}$ class, showing thermodynamic instability in both low- and high-temperature regimes. The introduction of rotation fundamentally modifies the topological characteristics: the Kerr black hole transitions to the $W^{0+}$ class, allowing coexistence of a stable small and an unstable large black hole at low temperatures, while no stable configurations persist at high temperatures. The presence of a quintessence field does not alter this universal classification.

For the AdS counterparts, the thermodynamic topology becomes markedly distinct. The Schwarzschild–AdS black hole corresponds to the $W^{0-}$ class, featuring the emergence of an unstable small and a stable large black hole only at high temperatures, with no black hole solution at low temperatures. In contrast, the Kerr–AdS configuration is classified as $W^{1+}$, where both small and large black holes can maintain stability across temperature ranges. This distinction highlights the significant influence of the AdS curvature and rotational degrees of freedom on the thermodynamic topology.

Furthermore, the inclusion of higher-order curvature corrections through the Gauss–Bonnet term in both 4D and 5D spacetimes preserves the $W^{0+}$ classification. These cases exhibit similar thermodynamic structures to the Kerr black hole in quintessence, characterized by the coexistence of stable and unstable branches at low temperatures and the disappearance of black hole states at high temperatures.

Overall, the comparative analysis demonstrates that the topological classification of black holes is sensitive to intrinsic parameters such as rotation and higher-curvature couplings, while the quintessence field mainly shifts the temperature-dependent stability ranges without altering the topological class itself. This work thus provides a new perspective for understanding the global stability structures of black holes in dark energy backgrounds.

\section*{Acknowledgments}
\hspace{0.5cm}
We are very grateful to Dr. P. K. Yerra for helpful discussion. We also acknowledge the anonymous referees for their valuable comments on improving our paper. The research was partially supported by the National Natural Science Foundation of China (Grant No.12265007). Also, the research of H.H. was supported by the Q-CAYLE project, funded by the European Union-Next Generation UE/MICIU/Plan de Recuperacion, Transformacion y Resiliencia/Junta de Castilla y Leon (PRTRC17.11), and also by project PID2023-148409NB-I00, funded by MICIU/AEI/10.13039/501100011033. Financial support of the Department of Education of the Junta de Castilla y Leon and FEDER Funds is also gratefully acknowledged (Reference: CLU-2023-1-05). Additionally, H. H. is grateful to Excellence project FoS UHK 2203/2025-2026 for the financial support.

\end{document}